\documentclass[superscriptaddress,amsmath,amssymb,pra,twocolumn]{revtex4-1}

\usepackage{graphicx}
\usepackage[dvipsnames]{xcolor}
\usepackage{hyperref}
\usepackage{braket,multirow}
\hypersetup{colorlinks = true, linkcolor = [rgb]{0.19411,0.51882,0.667058}, urlcolor = [rgb]{0.125490,0.29542,0.1647058}, citecolor = [rgb]{0.75882,0.37411,0.14117}}

\begin{document}

\title{Calculating Concentratable Entanglement in Graph States}

\author{Alice R. Cullen}
\author{Pieter Kok}
\affiliation{Department of Physics and Astronomy, The University of Sheffield, Hounsfield Road, S3 7RH Sheffield, United Kingdom}

\begin{abstract}
We propose a method to calculate the purity of reduced states of graph states entirely within the stabilizer formalism, using only the stabilizer generators for a given state. We apply this method to find the Concentratable Entanglement of graph states, with examples including a single qubit within any graph state, states proposed for use in quantum repeaters and states that maximise and minimise the overall Concentratable Entanglement.

\end{abstract}

\date{\today}

\maketitle

\section{Introduction}

The quantum phenomenon of entanglement \cite{einsteinCanQuantumMechanicalDescription1935, schrodingerDiscussionProbabilityRelations1935} that is present in quantum states associated with graphs \cite{heinMultipartyEntanglementGraph2004} makes such states a useful resource in quantum information processing. When used for Measurement Based Quantum Computation (MBQC) \cite{briegelMeasurementBased2009} and the One-Way Quantum Computer \cite{raussendorfOneWayQuantumComputer2001, raussendorfMeasurementbasedQuantumComputation2003}, the specific computation dictates the structure of the graph state. Conversely, the entanglement structure of a graph state determines its effectiveness in applications such as the quantum repeater \cite{briegelQuantumRepeatersRole1998, zwergerMeasurementbasedQuantumRepeaters2012, azumaAllphotonicQuantumRepeaters2015}, quantum error correction (QEC) codes \cite{gottesmanStabilizerCodesQuantum1997, schlingemannQuantumErrorCorrecting2001} and quantum secret sharing \cite{markhamGraphStatesQuantum2008}. An entanglement measure is required to investigate the structure of the multipartite entanglement within a graph state. 

Entanglement across the two subsystems within a bipartite quantum state is well understood. An initial qualitative test of entanglement is the Peres-Horodecki criterion \cite{peresSeparabilityCriterionDensity1996, horodeckiSeparabilityMixedStates1996}. A widely used quantitative measure is entanglement of formation \cite{bennettMixedstateEntanglementQuantum1996}, which for pure states coincides with the von Neumann entropy of the reduced state of either subsystem. The definition can be extended to mixed states by considering ensembles of pure states. Entanglement of formation for pure states can also be defined as a function of concurrence \cite{hillEntanglementPairQuantum1997}; a different entanglement measure that applies to pure states of two qubits, which can be adapted to apply to bipartite systems \cite{rungtaUniversalStateInversion2001}. 

Entanglement in multipartite states is more difficult to characterise and quantify, as the ways in which a state can be entangled increases with the number of parties involved. A widely cited example of different kinds of entanglement, first discussed by D\"{u}r et al.\ \cite{durThreeQubitsCan2000}, is demonstrated when considering three party entanglement, as the entanglement present in a three qubit GHZ state differs from that in a three qubit W state. It is not possible to transform either state to the other using Local Operations and Classical Communication (LOCC). Several existing multipartite measures are functions of bipartite entanglement measures \cite{meyerGlobalEntanglementMultiparticle2002, brennenObservableMeasureEntanglement2003, carvalhoDecoherenceMultipartiteEntanglement2004}, sometimes considered over multiple bipartitions of the state; for example this approach is applied to generalise concurrence to multipartite states \cite{carvalhoDecoherenceMultipartiteEntanglement2004}.

A recently proposed multipartite entanglement measure is Concentratable Entanglement \cite{beckeyComputableOperationallyMeaningful2021}. When defined for different cuts of qubits within a state, it relates to several existing measures \cite{meyerGlobalEntanglementMultiparticle2002, brennenObservableMeasureEntanglement2003, hillEntanglementPairQuantum1997, rungtaUniversalStateInversion2001, carvalhoDecoherenceMultipartiteEntanglement2004, fouldsControlledSWAPTest2021, wongPotentialMultiparticleEntanglement2001}, making it a candidate for adoption as the universal multipartite entanglement measure of pure states. Like other measures, one approach to calculating Concentratable Entanglement is to consider bipartitions of the multipartite quantum state. The purities of the corresponding reduced density matrices are required to evaluate the Concentratable Entanglement.

The multipartite entanglement of graph states has previously been investigated in terms of Schmidt measure and Schmidt rank \cite{heinMultipartyEntanglementGraph2004, heinEntanglementGraphStates2006}, which can be challenging to compute. Concentratable Entanglement offers a new measure that can be applied to graph states. Using the stabilizer formalism \cite{gottesmanStabilizerCodesQuantum1997}, which gives a convenient and efficient mathematical description of graph states, we have developed a method reliant only on the stabilizer generators for determining reduced state purities and thus Concentratable Entanglement for graph states. This allows for calculation of the Concentratable Entanglement for examples such as single qubits in graph states, for subsets of qubits within graph states proposed for use in quantum repeaters and as a measure of the overall entanglement for any graph state.

\section{Background}
We begin by outlining the mathematical background required to define Concentratable Entanglement \cite{beckeyComputableOperationallyMeaningful2021}, graph states within the stabilizer formalism and how measurements of individual qubits affect a graph state \cite{heinMultipartyEntanglementGraph2004}.

\subsection{Concentratable Entanglement}
Concentratable Entanglement \cite{beckeyComputableOperationallyMeaningful2021} can be used as a measure of multipartite entanglement of a pure quantum state $\ket{\psi}$. For a state of $n$ qubits with labels in $\mathcal{S}=\{1,\ldots,n\}$, the Concentratable Entanglement can be calculated for the full set to determine the overall entanglement, or it can be applied to any non-empty subset $s \subseteq \mathcal{S}$ to investigate the structure of the entanglement within the state. Operationally, the Concentratable Entanglement corresponds to the probability of producing Bell pairs in a SWAP test \cite{buhrmanQuantumFingerprinting2001}. The Concentratable Entanglement of $\ket{\psi}$ can be defined in terms of SWAP test outcomes, making it possible to efficiently calculate its value using quantum computers and two copies of $\ket{\psi}$.

However, without the physical resource of a quantum computer, the definition of the Concentratable Entanglement in terms of purities still offers a method to determine the entanglement of a (subset of a) state. For a non-empty set of qubit labels $s \subseteq \mathcal{S}$, the Concentratable Entanglement is \cite{beckeyComputableOperationallyMeaningful2021}
\begin{equation}
\label{CE}
C_{\ket{\psi}}(s) = 1 - \frac{1}{2^{|s|}} \sum_{\alpha \in \mathcal{P}(s)} \text{Tr} \rho_{\alpha}^{2},
\end{equation}
where $\rho_{\alpha}$ is the density matrix of the reduced state with qubits labelled by the set $\alpha$, which is a subset of the power set $\mathcal{P}(s)$. This requires the calculation of up to $2^n$ purities, including the trivial case $\text{Tr}\rho_{\emptyset}^2 = 1$.

\subsection{Graph States}
In order to describe graph states, we first define the Pauli matrices as
\begin{equation}
X = \begin{pmatrix}
0 & 1\\
1 & 0
\end{pmatrix}, \quad
Y = \begin{pmatrix}
0 & -i\\
i & 0
\end{pmatrix}, \quad
Z = \begin{pmatrix}
1 & 0\\
0 & -1
\end{pmatrix}.
\end{equation}
A graph $G=(V,E)$ is determined by a set of vertices $V$ and a set of edges $E$ connecting pairs of vertices. Graphs can be used as pictorial representations of entangled quantum states. Each vertex represents a qubit in the $\ket{+}$ eigenstate, i.e.,\ the eigenstate of the Pauli $X$ operator corresponding to the eigenvalue $+1$. Edges denote interactions between qubits in the form of controlled-$Z$ (CZ) operations. Such quantum states are known as graph states. Since there is no entanglement between disconnected graphs, we consider only connected graphs.

The neighbourhood $n(a)$ of a vertex $a \in V$ is the set of vertices connected to $a$ by a single edge. In a graph state this identifies the qubits that jointly undergo a CZ operation with $a$.

Each qubit in the graph state $\ket{G}$ can be assigned a Hermitian operator of the form \cite{raussendorfOneWayQuantumComputer2001}
\begin{equation}
\label{SG}
S_a = X_a \prod_{b \in n(a)} Z_b.
\end{equation}
These $n$ operators form the generators of an Abelian group, known as the stabilizer $\mathfrak{S}_G$ of the state $\ket{G}$. The $2^n$ elements in the stabilizer are the products of all possible subsets of the stabilizer generators including the identity $I$. The stabilizer is a subgroup of the Pauli group for $n$ qubits, i.e.,\ $\mathfrak{S}_G \subset \mathcal{G}_n$.  The state $\ket{G}$ is an eigenstate with eigenvalue $+1$ for any element $S$ of the stabilizer
\begin{equation}
\label{SP}
S\ket{G} = \ket{G} \quad \forall S \in \mathfrak{S}_G.
\end{equation}
The stabilizer formalism provides an efficient way to fully specify graph states.

\subsection{Measurement}
When a gate from the Clifford group or a Pauli measurement is applied to a stabilizer state, the resulting state is also a stabilizer state. This is a useful property that can be applied when considering measurements on qubits within graph states.

The simplest measurement to describe on a graph state is that of the Pauli $Z$ operator on a single qubit. As an immediate consequence of such a measurement, the measured qubit is removed from the graph state. Depending on the measurement outcome with possible values $+1$ or $-1$, the individual qubit is in one of the two eigenstates of the $Z$ operator, $\ket{0}$ or $\ket{1}$, with stabilizer generator $Z$ or $-Z$ respectively. 

Though the measured qubit is disconnected from the graph state, it still affects the remaining state. When qubit $a \in V$ is measured, the graph becomes $G'=G-\{a\}$ where all edges to vertex $a$ are removed. The state of the remaining qubits is $U\ket{G-\{a\}}$ where $U$ is a unitary operator determined by the measurement outcome of $a$. For the outcomes $\pm1$ the respective unitaries are
\begin{equation}
\label{Unit}
U_+=I \quad \text{and} \quad U_- = \prod_{b \in n(a)} Z_b.
\end{equation}
In the stabilizer formalism the stabilizer generator $S_b'$ for each remaining vertex $b \in V'$ will again take the form of Eq.\ (\ref{SG}). The unitary operator corresponding to the outcome is applied to each generator as $US_b'U^{\dagger}$ to give the generators for the state $U\ket{G-\{a\}}$.

An alternative method is to take the original set of stabilizer generators for $\ket{G}$ and set the stabilizer generator for the measured qubit to $\pm Z_a$. The generators for qubits that were not within the neighbourhood of qubit $a$ remain unchanged. However the generators for qubits within the neighbourhood of $a$, which contain the term $Z_a$, are multiplied by the new generator $\pm Z_a$, allowed by the group structure of the new stabilizer. This produces the new stabilizer generators without explicitly drawing the graph representing the quantum state.

\section{Purities of Reduced Graph States}

Calculating the purities required within the Concentratable Entanglement relies on the reduced states of different subsets of qubits. Within the stabilizer formalism, a reduced state can be described as a mixture of several pure states, each of which has its own set of stabilizer generators. We show that the number of distinct sets of stabilizer generators corresponds to the purity of the reduced state that the sets of generators collectively describe.

Consider a bipartition $(A,B)$ of qubits within the graph $G$. The reduced state of the qubits in $B$ is found by tracing out the qubits in $A$
\begin{equation}
\rho_B = \text{Tr}_A(\ket{G}\bra{G}).
\end{equation}
This partial trace is equivalent to measuring the qubits in $A$ and discarding the outcome. These measurements can be performed in the $Z$-direction, where the outcomes $\pm 1$ each occur with probability $1/2$.

Through repeated application of the $Z$ measurement rule and unitaries given in Eq.\ (\ref{Unit}), and considering all possible outcomes, the density operator for the reduced state of the qubits in $B$ is
\begin{equation}
\label{PT}
\rho_B = \frac{1}{2^{|A|}} \sum_{\textbf{z} \in \mathbb{F}_2^A} U(\textbf{z}) \ket{G-A}\bra{G-A} U(\textbf{z})^{\dagger},
\end{equation}
where $\mathbb{F}_2$ is the finite field of two elements $\{0,1\}$. An element of the finite field $\mathbb{F}_2^A$ is a bitstring $\textbf{z}$ of length $|A|$. In $\mathbb{F}_2^A$ addition and subtraction are equivalent and performed modulo 2 without carry. The local unitaries are \cite{heinMultipartyEntanglementGraph2004}
\begin{equation}
\label{Unitary}
U(\textbf{z}) = \prod_{a \in A} \Bigg( \prod_{b \in n(a) \cap B}Z_b \Bigg) ^{z_a},
\end{equation}
where $z_a = 0$ represents the outcome $+1$ and $z_a = 1$ for $-1$. Within the unitaries, only qubits in the neighbourhood of measured qubits which are also in $B$ need to be considered. This is because the individual qubits in $A$ can be measured in any order, making it possible for neighbourhood qubits to already be disconnected from the graph. Further, $Z_b$ for $b \in A$ has no effect on the state $\ket{G-A}$, which does not contain qubits in $A$.

For completeness, we prove the following properties of these unitary operators that have previously been stated and applied \cite{heinMultipartyEntanglementGraph2004}, as they provide the basis for proving our own result regarding the purities of reduced states.

\textit{Lemma}. (Hein-Eisert-Briegel) 1.\ For different measurement outcomes $\textbf{z}$ and $\textbf{z}'$ to give the same unitary operator, i.e. $U(\textbf{z}) =  U(\textbf{z}')$, the bitstrings must satisfy \cite{heinMultipartyEntanglementGraph2004}
\begin{equation}
U(\textbf{z}-\textbf{z}') = I.
\end{equation}

2.\ If $U(\textbf{z}) \neq U(\textbf{z}')$ then the states $U(\textbf{z})\ket{G-A}$ and $U(\textbf{z}')\ket{G-A}$ are orthogonal.

\textit{Proof}. 1.\ Using the definition of the unitary operator in Eq.\ (\ref{Unitary}),
\begin{equation}
\label{Uzz}
\begin{split}
U(\textbf{z}-\textbf{z}') &= \prod_{a \in A} \Bigg( \prod_{b \in n(a) \cap B}Z_b \Bigg) ^{z_a-z'_a}\\
&= \prod_{a \in A} \Bigg( \prod_{b \in n(a) \cap B}Z_b \Bigg) ^{z_a} \prod_{a' \in A} \Bigg( \prod_{b' \in n(a') \cap B}Z_{b'} \Bigg) ^{-z'_{a'}}\\
&= \prod_{a \in A} \Bigg( \prod_{b \in n(a) \cap B}Z_b \Bigg) ^{z_a} \prod_{a' \in A} \Bigg( \prod_{b' \in n(a') \cap B}Z_{b'}^{-1} \Bigg) ^{z'_{a'}}\\
&=U(\textbf{z})U(\textbf{z}')^{\dagger}.
\end{split}
\end{equation}
This is equal to the identity
\begin{equation}
U(\textbf{z}-\textbf{z}') = U(\textbf{z})U(\textbf{z}')^{\dagger} = I\\
\end{equation}
and by right multiplying by $U(\textbf{z}')$, this gives $U(\textbf{z}) = U(\textbf{z}')$.

2.\ If $U(\textbf{z}) \neq U(\textbf{z}')$ then $U(\textbf{z})U(\textbf{z}')^{\dagger}  \neq I$ using the above proof. $U(\textbf{z})U(\textbf{z}')^{\dagger}$ is a product of Pauli $Z$ operators, as shown in Eq.\ (\ref{Uzz}), which in the simplest case consists of a single operator $Z_j$ where $j \in B$. Qubit $j$ in the remaining graph $G'=G-A$ is associated with a stabilizer generator $S'_j$ in the form of Eq.\ (\ref{SG}) that satisfies $S'_j\ket{G'} = \ket{G'}$. The inner product of the states $U(\textbf{z})\ket{G-A}$ and $U(\textbf{z}')\ket{G-A}$ is
\begin{equation}
\label{Orth}
\begin{split}
\bra{G'}U(\textbf{z}')^{\dagger}U(\textbf{z})\ket{G'} &= \bra{G'}Z_j \ket{G'}\\
&= \bra{G'} Z_j S'_j\ket{G'}\\
&= \bra{G'}Z_j X_j \textstyle{\prod_{b \in n(j)}}Z_b \ket{G'}\\
&= \bra{G'}(-X_j Z_j) \textstyle{\prod_{b \in n(j)}}Z_b \ket{G'}\\
&= - \bra{G'}(X_j \textstyle{\prod_{b \in n(j)}}Z_b)Z_j  \ket{G'}\\
&= - \bra{G'}S_j^{\prime \dagger} Z_j\ket{G'}\\
&= - \bra{G'}Z_j\ket{G'}\\
&= - \bra{G'}U(\textbf{z}')^{\dagger}U(\textbf{z})\ket{G'}.
\end{split}
\end{equation}
This is satisfied only when $\bra{G'}U(\textbf{z}')^{\dagger}U(\textbf{z})\ket{G'} = 0$ so the states $U(\textbf{z})\ket{G-A}$ and $U(\textbf{z}')\ket{G-A}$ are orthogonal.

More generally $U(\textbf{z})U(\textbf{z}')^{\dagger} $ can be written as
\begin{equation}
U(\textbf{z})U(\textbf{z}')^{\dagger}  = Z_j \prod_{c \in C} Z_c
\end{equation}
where $C$ is a subset of $B \setminus j$. $\textstyle{\prod_{c \in C} Z_c}$ commutes with $S'_j$ so replacing $Z_j$ by the more general expression for $U(\textbf{z})U(\textbf{z}')^{\dagger}$ in Eq.\ (\ref{Orth}), the states $U(\textbf{z})\ket{G-A}$ and $U(\textbf{z}')\ket{G-A}$ satisfy 
$\bra{G-A}U(\textbf{z}')^{\dagger}U(\textbf{z})\ket{G-A} = 0$ and hence are orthogonal. \hfill $\square$

The unitary operators $U(\textbf{z})$ can used within the stabilizer formalism to find the reduced state of qubits in $B$. The set of stabilizer generators for each outcome must be considered separately. The remaining graph $G-A$ can be found by removing all edges to qubits in $A$ and the stabilizer generator $S'_b$ for each qubit $b \in B$ is given by Eq.\ (\ref{SG}). Then for each outcome, the corresponding unitary operator must be applied to each stabilizer generator according to $U(\textbf{z})S'_bU(\textbf{z})^{\dagger}$, giving a new stabilizer for each unitary operator. 

An alternative method does not require calculation of the unitary operators corresponding to each measurement outcome. Instead, in the set of stabilizer generators for the original graph $G$, the stabilizer generator for each qubit $a \in A$ can be set to $\pm Z_a$. Then the group structure of the stabilizer can be used to determine the stabilizer generators for the remaining graph only in terms of qubits in $B$. This process is repeated to give a set of stabilizer generators for each measurement outcome.

The following theorem provides a link between the sets of stabilizer generators that describe a reduced state and the purity of that reduced state. This is the main component in calculating the Concentratable Entanglement.

\textit{Theorem 1}. Consider a graph state bipartitioned into sets $A$ and $B$. Let $k \in \mathbb{N}$ denote the number of distinct sets of stabilizer generators for the different measurement outcomes when measuring the qubits in $A$. The purity of the reduced state of qubits in $B$ is then
\begin{equation}
\label{Theorem}
\text{Tr} \rho_B ^2 = \frac{1}{k}.
\end{equation}

\textit{Proof}. When measuring the qubits in $A$ there are $2^{|A|}$ possible outcomes. Each outcome is represented by a bitstring $\textbf{z} \in \mathbb{F}_2^A$, which is associated with a unitary operator $U(\textbf{z})$. Therefore there are $2^{|A|}$ unitary operators. However, these unitaries need not be unique.

For any graph state and subset $A$ of qubits, the outcome represented by the bitstring made entirely of zeros leads to the identity operator
\begin{equation}
U(\textbf{0}) = I.
\end{equation}
If $U(\textbf{0})$ is the sole identity operator, operators $U(\textbf{z})$ and $U(\textbf{z}')$ are equal when $\textbf{z}-\textbf{z}'=\textbf{0}$ using the Hein-Eisert-Briegel Lemma. This is solved trivially by $\textbf{z}=\textbf{z}'$ but has no further solutions. Therefore when one operator is the identity, all remaining unitaries take distinct values.

Consequently, for any unitary to occur more than once, there must be further operators equal to the identity. If a second operator is the identity, i.e., $U(\textbf{w}) = I$ for some $\textbf{w} \neq \textbf{0}$, the remaining elements in the finite field can be matched into pairs whose difference modulo 2 is $\textbf{w}$. Therefore there are $2^{|A|}/2 = 2^{|A|-1}$ different unitaries, each of which occurs twice.

If a third unitary is set to the identity, $U(\textbf{v}) = I$ for $\textbf{v} \notin \{\textbf{0},\textbf{w}\},$ then it follows that a fourth unitary, $U(\textbf{v}')$ where $\textbf{v}' = \textbf{v} - \textbf{w}$, must also be the identity. Hence
\begin{equation}
U(\textbf{0}) = U(\textbf{w}) = U(\textbf{v}) = U(\textbf{v}') = I.
\end{equation}
For any remaining $\textbf{z} \in \mathbb{F}_2^A$, $U(\textbf{z})$ is equal to three other unitaries, those for which the bitstring differs from $\textbf{z}$ by $\textbf{w}$, $\textbf{v}$ or $\textbf{v}'$. There are $2^{|A|}/4 = 2^{|A|-2}$ different unitary operators, each occurring 4 times.

More generally, if there are $2^j$ unitary operators that are the identity where $0 \leq j \leq |A|$, then there are $k=2^{|A|-j}$ different values for the unitary operators, each of which has multiplicity $2^j$.

To find the purity of the reduced state, first the density operator must be squared. Using Eq.\ (\ref{PT}), this gives
\begin{equation}
\rho_B^2 = \frac{1}{2^{2|A|}}
\sum_{\textbf{z}, \textbf{z}'} U(\textbf{z}) \ket{G'} \bra{G'} U(\textbf{z})^{\dagger} U(\textbf{z}') \ket{G'} \bra{G'} U(\textbf{z}')^{\dagger},
\end{equation}
where $G'=G-A$ for brevity. By the Hein-Eisert-Briegel Lemma, if $U(\textbf{z}) \neq U(\textbf{z}')$ then $\bra{G'} U(\textbf{z})^{\dagger} U(\textbf{z}') \ket{G'} = 0$ since the states $U(\textbf{z}) \ket{G'}$ and $U(\textbf{z}') \ket{G'}$ are orthogonal. In the $2^j$ cases when $U(\textbf{z}) = U(\textbf{z}')$, including the case $\textbf{z} = \textbf{z}'$, $\bra{G'} U(\textbf{z})^{\dagger} U(\textbf{z}') \ket{G'} = 1$. This leads to
\begin{equation}
\rho_B^2 = \frac{2^j}{2^{2|A|}} \sum_{\textbf{z} \in \mathbb{F}_2^A} U(\textbf{z}) \ket{G'}\bra{G'} U(\textbf{z})^{\dagger} =\frac{1}{k} \rho_B.
\end{equation}

The trace is then taken, resulting in a purity given by
\begin{equation}
\text{Tr}\rho_B^2 = \frac{1}{k},
\end{equation}
since $\text{Tr}\rho_B = 1$.

If there are $k$ different unitaries, this leads to $k$ different sets of stabilizer generators to describe the graph state once the qubits in $A$ have been removed by measurement. Each of the $k$ unitaries is applied to give a set of generators according to $U S'_b U^{\dagger}$ for all $b \in B$. \hfill $\square$

For any bipartition, the purities of the reduced states $\rho_A$ and $\rho_B$ are related by \cite{nielsenQuantumComputationQuantum2010}
\begin{equation}
\label{SD}
\text{Tr}\rho_A^2 = \text{Tr}\rho_B^2,
\end{equation}
due to the Schmidt decomposition theorem. When calculating purities using Theorem 1, the smaller partition from $(A,B)$ can be traced out, reducing the number of bipartitions of an $n$ qubit state to consider by one half. 

Theorem 1 also applies when using the alternative method of finding sets of stabilizer generators without calculating the unitary operators, since this method results in the same sets of stabilizer generators.

\section{Examples}
The result of Theorem 1 can be applied within the stabilizer formalism to calculate the Concentratable Entanglement for cuts of qubits of graph states with different designs and applications. We will consider several examples here.

\textbf{1. Single qubits.} The Concentratable Entanglement for a set containing a single qubit $a$ in an $n$ qubit graph state $\ket{G}$ is
\begin{equation}
C_{\ket{G}}(\{a\}) = \frac{1}{4}.
\end{equation}
To show this, consider any connected graph state of $n$ qubits. To find the Concentratable Entanglement for a set $s$ containing a single qubit, the purity of the reduced state of the single qubit must be calculated. The reduced state can be found by tracing out the other $n-1$ qubits and considering all $2^{n-1}$ possible measurement outcomes. However this can be simplified by instead using Eq.\ (\ref{SD}) and only tracing out the qubit of interest. When bipartioning the set $\mathcal{S} = \{1,\ldots,n\}$ of all qubit labels, the set $A = \{a\}$ contains a single qubit and $B = \mathcal{S} \setminus \{a\}$ contains the remaining $n-1$ qubits.

The stabilizer generators for the graph take three forms. Firstly, qubit $a$ has the stabilizer generator
\begin{equation}
S_a = X_a \prod_{b \in n(a)} Z_b.
\end{equation}
Second, the stabilizer generator for each qubit $b \in n(a)$ is
\begin{equation}
S_b = Z_a X_b \prod_{k \in \tilde{n}(a)} Z_{k},
\end{equation}
where $\tilde{n}(a) = n(b) \setminus \{a\}$.
Finally, qubits $c \in B \setminus n(a) $ have stabilizer generators of the form
\begin{equation}
S_c = X_c \prod_{m \in n(c)} Z_{m},
\end{equation}
where $n(c)$ cannot contain qubit $a$.

By tracing out qubit $a$, its stabilizer generator must be updated to $\pm Z_a$. In both cases generators in the form of $S_c$ remain valid. However each $S_b$ must be multiplied by $\pm Z_a$ so the new generator contains only Pauli operators acting on qubits in $B$. The possible sets of stabilizer generators are\\

\begin{minipage}{0.2\textwidth}
\centering $Z_a$
\begin{equation*}
\begin{split}
S_b' & = X_b \prod_{k \in \tilde{n}(b)} Z_{k}\\
S_c' & = X_c \prod_{m \in n(c)} Z_{m}
\end{split}
\end{equation*}
\end{minipage}
\begin{minipage}{0.22\textwidth}
\centering $-Z_a$
\begin{equation*}
\begin{split}
S_b' & = - X_b \prod_{k \in \tilde{n}(b)} Z_{k}\\
S_c' & = \phantom{-} X_c \prod_{m \in n(c)} Z_{m},
\end{split}
\end{equation*}
\end{minipage}
where there is a stabilizer generator in both sets for each $b \in n(a)$ and each $c \in B \setminus n(a)$. The stabilizer generators for qubits in the neighbourhood of qubit $a$ acquire a minus sign when the measurement outcome is $-1$.

There are two possible sets of stabilizer generators to describe the reduced states of qubits in $B$. Applying Theorem 1 and Eq.\ (\ref{SD}), the purity is
\begin{equation}
\text{Tr}\rho_B^2 = \frac{1}{2} = \text{Tr}\rho_a^2.
\end{equation}
Therefore the Concentratable Entanglement for a single qubit within any connected graph state, where $s = \{a\}$, is
\begin{equation}
C_{\ket{G}}(\{a\}) = 1 - \frac{1}{2}(1 + \text{Tr}\rho_a^2) = \frac{1}{4}.
\end{equation}

\begin{figure}[t]
\begin{center}
\includegraphics[width=0.4\textwidth]{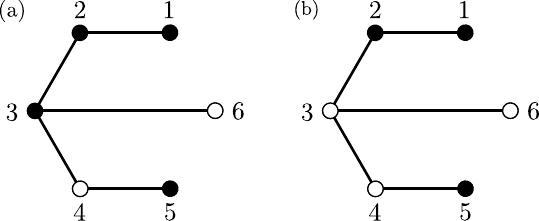}
\caption{A six qubit graph state. The qubit labels can be bipartitioned, with qubits in set $A$ represented by hollow vertices and qubits in set $B$ by filled vertices. (a) $A = \{4,6\}$, $B = \{1,2,3,5\}$. (b) $A = \{3,4,6\}$, $B = \{1,2,5\}$.}
\label{6qubits}
\end{center}
\end{figure}

\textbf{2. A six qubit graph.} Fig.\ \ref{6qubits} shows a graph state with qubit labels $\mathcal{S}= \{1, \ldots, 6\}$. This is graph No.\ 13 according to the standard numbering of graph state local unitary (LU) equivalence classes \cite{heinMultipartyEntanglementGraph2004, danielsenClassificationAllSelfDual2006, cabelloOptimalPreparationGraph2011}. The graph is described by stabilizer generators
\begin{equation}
\begin{split}
S_1 &= X_1 Z_2\\
S_2 &= Z_1 X_2 Z_3\\
S_3 &= \phantom{X_1} Z_2 X_3 Z_4 \phantom{X_5} Z_6\\
S_4 &= \phantom{X_1 Z_2} Z_3 X_4 Z_5\\
S_5 &= \phantom{X_1 X_2 Z_3} Z_4 X_5\\
S_6 &= \phantom{X_1 Z_2} Z_3 \phantom{X_4 Z_5} X_6.
\end{split}
\end{equation}

To find the purity of the reduced state of qubits 1, 2, 3 and 5, qubits 4 and 6 must be traced out. The stabilizer generators for qubits 4 and 6 are set to $\pm Z_4$ and $\pm Z_6$ to account for all possible measurement outcomes, where qubits 4 and 6 collapse into the eigenstates of the $Z$ operator. This results in the following sets of stabilizer generators to describe the remaining state\\

\begin{minipage}{0.225\textwidth}
\centering $\{Z_4,Z_6\}$
\begin{equation*}
\begin{split}
S_1' & = X_1 Z_2 \phantom{Z_3}\\
S_2' & = Z_1 X_2 Z_3\\
S_3' & = \phantom{Z_1} Z_2 X_3\\
S_5' & = \phantom{X_1 Z_2 Z_3} X_5
\end{split}
\end{equation*}
\end{minipage}
\begin{minipage}{0.225\textwidth}
\centering $\{Z_4,-Z_6\}$
\begin{equation*}
\begin{split}
S_1' & = X_1 Z_2 \phantom{Z_3}\\
S_2' & = Z_1 X_2 Z_3\\
S_3' & = \phantom{z} {-Z_2} X_3\\
S_5' & = \phantom{X_1 Z_2 Z_3} X_5
\end{split}
\end{equation*}
\end{minipage}\\

\begin{minipage}{0.225\textwidth}
\vspace{10pt}
\centering $\{-Z_4,Z_6\}$
\begin{equation*}
\begin{split}
S_1' & = X_1 Z_2 \phantom{Z_3}\\
S_2' & = Z_1 X_2 Z_3\\
S_3' & = \phantom{z} {-Z_2} X_3\\
S_5' & = \phantom{X_1 Z_2 z} {-}X_5
\end{split}
\end{equation*}
\end{minipage}
\begin{minipage}{0.225\textwidth}
\vspace{10pt}
\centering $\{-Z_4,-Z_6\}$
\begin{equation*}
\begin{split}
S_1' & = X_1 Z_2 \phantom{Z_3}\\
S_2' & = Z_1 X_2 Z_3\\
S_3' & = \phantom{Z_1} Z_2 X_3\\
S_5' & = \phantom{X_1 Z_2 z} {-}X_5.
\end{split}
\end{equation*}
\end{minipage}\\
There are four distinct sets of stabilizer generators and therefore by Theorem 1 the purity of the reduced state of qubits 1, 2, 3 and 5 is
\begin{equation}
\text{Tr}\rho_{1235}^2 = \frac{1}{4}.
\end{equation}

Instead, if qubits 3, 4 and 6 are traced out, the stabilizer generators for each of these qubits must be changed to consider all possible outcomes. Since the stabilizer generators where qubits 4 and 6 are traced out have already been calculated, only the stabilizer generator for qubit 3 needs to be updated. The stabilizer generator for qubit 3 is set to $\pm Z_3$ within each set. Therefore the sets of stabilizer generators for the remaining qubits are\\

\begin{minipage}{0.225\textwidth}
\centering $\{Z_3,Z_4,Z_6\}$, $\{Z_3,Z_4,-Z_6\}$
\begin{equation*}
\begin{split}
S_1'' & = X_1 Z_2\\
S_2'' & = Z_1 X_2\\
S_5'' & = \phantom{X_1 Z_2} X_5
\end{split}
\end{equation*}
\end{minipage}
\begin{minipage}{0.225\textwidth}
\centering $\{-Z_3,Z_4,Z_6\}$, $\{-Z_3,Z_4,-Z_6\}$
\begin{equation*}
\begin{split}
S_1'' & = \phantom{-} X_1 Z_2\\
S_2'' & = -Z_1 X_2\\
S_5'' & = \phantom{-X_1 Z_2} X_5
\end{split}
\end{equation*}
\end{minipage}\\

\begin{minipage}{0.225\textwidth}
\vspace{10pt}
\centering $\{Z_3,-Z_4,Z_6\}$, $\{Z_3,-Z_4,-Z_6\}$
\begin{equation*}
\begin{split}
S_1'' & = X_1 Z_2\\
S_2'' & = Z_1 X_2\\
S_5'' & = \phantom{-Z} {-X_5}
\end{split}
\end{equation*}
\end{minipage}
\begin{minipage}{0.225\textwidth}
\vspace{10pt}
\centering $\{-Z_3,-Z_4,Z_6\}$, $\{-Z_3,-Z_4,-Z_6\}$
\begin{equation*}
\begin{split}
S_1'' & = \phantom{-} X_1 Z_2\\
S_2'' & = -Z_1 X_2\\
S_5'' & = \phantom{-{-Z}} {-X_5}.
\end{split}
\end{equation*}
\end{minipage}\\

Though there are eight possible measurement outcomes, this results in only four distinct sets of stabilizer generators. Therefore by Theorem 1 the purity of the reduced state of qubits 1, 2 and 5 is
\begin{equation}
\text{Tr}\rho_{125}^2 = \frac{1}{4}.
\end{equation}
This is also the purity of the state $\rho_{346}$ due to Eq.\ (\ref{SD}) and of the states $\rho_{145}$ and $\rho_{236}$ because of the symmetry of the graph in Fig.\ \ref{6qubits}.

This method can be used to find the purities of all possible reduced states of the graph state. For the six bipartitions where the smaller set contains a single qubit, the purity is $1/2$ from Example 1. When the smaller set in the bipartition contains two qubits, a purity of $1/4$ occurs twelve times and a purity of $1/2$ occurs three times. For the ten bipartitions where each set has three elements, the purity is $1/8$ in four cases, $1/4$ in four cases and $1/2$ in two cases. These values can be used to calculate the Concentratable Entanglement of any subset of qubits and the overall Concentratable Entanglement of the graph state:
\begin{widetext}
\begin{equation}
\begin{split}
C_{\ket{G_{13}}}(\mathcal{S}) &= 1 -\frac{1}{2^6} \bigg[1 + 6 \cdot \frac{1}{2} + \bigg(12 \cdot \frac{1}{4} + 3 \cdot \frac{1}{2}\bigg) + \bigg(8 \cdot \frac{1}{8} + 8 \cdot \frac{1}{4} + 4 \cdot \frac{1}{2}\bigg) + \bigg(12 \cdot \frac{1}{4} + 3 \cdot \frac{1}{2}\bigg) + 6 \cdot \frac{1}{2} + 1 \bigg]\\
&= \frac{21}{32}.
\end{split}
\end{equation}
\end{widetext}

An alternative measure of entanglement with respect to a bipartition $(A,B)$ is the Schmidt rank, defined as \cite{heinEntanglementGraphStates2006}
\begin{equation}
\text{SR}_B(\psi) = \log_2[\text{rank}(\rho_B)].
\end{equation}
This measure is related to the purity of the reduced state of $B$ according to \cite{heinEntanglementGraphStates2006}
\begin{equation}
\label{SR}
\text{SR}_B(\psi) = -\log_2[\text{Tr}(\rho_B^2)].
\end{equation}
The rank index $RI_m$ considers all bipartitions where the smaller set contains $m$ qubits \cite{heinMultipartyEntanglementGraph2004}. It lists the number of times the Schmidt rank $m$, then $m-1$, through to $1$ occurs. For graph No.\ 13 in Fig \ref{6qubits}, $\text{RI}_2 = (12,3)$ and $\text{RI}_3 = (4,4,2)$: for smaller sets in the bipartition of qubit labels of the graph containing two qubits, there are twelve sets with a Schmidt rank of 2 and three sets with a Schmidt rank of 1. For smaller sets of three qubits, the Schmidt rank 3 occurs four times, 2 occurs four times and 1 occurs twice. This corresponds to the occurrences of different purities of reduced states calculated for the Concentratable Entanglement. Using Eq.\ (\ref{SR}), the rank index can be used to give the number of splits where the purity is $(2^{-m},\ldots,1)$ for bipartitions with a smaller set of $m$ qubits. For any connected graph of seven or fewer vertices, the Concentratable Entanglement for $s=\mathcal{S}$ can be calculated using the results for the rank index in Table $\mbox{II}$ of \cite{heinMultipartyEntanglementGraph2004}.

\begin{figure}[b!]
\begin{center}
\includegraphics[width=0.2\textwidth]{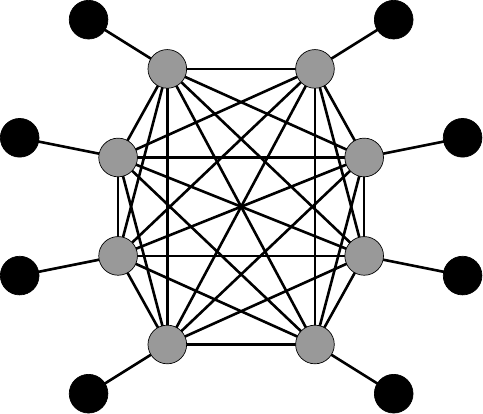}
\caption{A sixteen qubit `snowflake' state proposed for use in repeaters \cite{azumaAllphotonicQuantumRepeaters2015}. Grey circles represent 1st leaf qubits, each of which is connected to a 2nd leaf qubit, shown in black.}
\label{Photonic}
\end{center}
\end{figure}

\textbf{3. Graph states for repeaters.} Azuma et al.\ \cite{azumaAllphotonicQuantumRepeaters2015} propose a quantum repeater design based on graph states; repeater stations create `snowflake' graph states as shown in Fig.\ \ref{Photonic}, to perform entanglement swapping and set up direct entanglement over larger distances. The inner 1st leaf qubits (grey) form a fully connected graph so there is a direct path between any two qubits. Each 1st leaf qubit is further connected to a single 2nd leaf qubit (black).

Consider a snowflake state of $2n$ qubits where the 1st leaf qubits have labels in the set $\mathcal{S}_{1\text{st}} = \{1_1,2_1,\ldots,n_1\}$ and the 2nd leaf qubit labels are $\mathcal{S}_{2\text{nd}} = \{1_2,2_2,\ldots,n_2\}$. Qubit $a_2$ is the 2nd leaf qubit paired with 1st leaf qubit $a_1$. The stabilizer generator for each 1st leaf qubit $a_1 \in \mathcal{S}_1$ is
\begin{equation}
S_{a_1} = X_{a_1} Z_{a_2} \prod_{b \in \mathcal{S}_1 \setminus \{a_1\}} Z_b.
\end{equation}
Each 2nd leaf qubit $a_2 \in \mathcal{S}_2$ has a stabilizer generator of the form
\begin{equation}
S_{a_2} = X_{a_2} Z_{a_1}.
\end{equation}

When measuring a 1st leaf qubit in the $Z$ direction, the outcome $-1$ introduces a minus sign to the stabilizer generator for its 2nd leaf qubit partner and in the stabilizer generators for all other 1st leaf qubits, as they are all in the neighbourhood. For a 2nd leaf qubit, the measurement outcome $-1$ introduces a minus sign only to the stabilizer generator of its paired qubit in the 1st leaf. Therefore, if a pair $\{a_1, a_2\}$ is traced out, there are only two possible sets of stabilizer generators. The 1st leaf qubit stabilizer generators either all acquire a minus sign or they do not.

When $m \leq n$ qubits are traced out, where no two qubits are a pair, each measurement outcome results in a unique set of stabilizer generators for the remaining qubits. The purity of the reduced state left when tracing out these $m$ qubits is
$2^{-m}$. Using this result to calculate each of the purities required, the Concentratable Entanglement for sets $s$ of $n$ qubits containing no pairings, including the sets containing exclusively 1st leaf or exclusively 2nd leaf qubits, is
\begin{equation}
C_{\ket{G}}(s) = 1 - \frac{1}{2^n}\sum_{m=0}^n {n \choose m} \frac{1}{2^m} = 1 - \bigg(\frac{3}{4}\bigg)^n.
\end{equation}
This is the maximum Concentratable Entanglement possible for a set of $n$ qubits within a graph state of $2n$ qubits. Therefore within a snowflake state, all of the 1st leaf qubits or all of the 2nd leaf qubits exhibit maximal entanglement with respect to the rest of the state.

\textbf{4. Limits on overall entanglement.} Concentratable Entanglement provides a measure of the overall entanglement of a state when applied to the set $\mathcal{S}$. There are theoretical bounds on the maximum and minimum values that the Concentratable Entanglement can take.

The minimum value of Concentratable Entanglement occurs when the purity of each possible reduced state is maximised. Since a connected graph represents an entangled pure state, any reduced state cannot be pure, excluding the reduced state containing no qubits where $\text{Tr}\rho_{\emptyset}^2 = 1$. The maximum value each of these purities can take according to Theorem 1 is $1/2$, where there are two sets of stabilizer generators to describe the remaining qubits within the reduced state. The minimum value of Concentratable Entanglement for a graph state of $n$ qubits is therefore
\begin{equation}
\begin{split}
C_{\ket{G}}(\mathcal{S}) &= 1 - \frac{1}{2^n}\bigg(\text{Tr}\rho_{\emptyset}^2 + \text{Tr}\rho_{\mathcal{S}}^2 + \sum_{i=1}^{n-1}{n \choose i}\frac{1}{2}\bigg)\\
&= \frac{1}{2} - \frac{1}{2^n}.
\end{split}
\end{equation}
This is the Concentratable Entanglement of the GHZ state \cite{beckeyComputableOperationallyMeaningful2021}. The $n$ qubit GHZ state is LU equivalent to the star graph and the complete graph, which take this value for Concentratable Entanglement.

If instead the purity of each reduced state is minimised, the Concentratable Entanglement will be maximised. It is not always possible to minimise the purity of all reduced states due to the structure of entanglement and the dependent relationships between different subsets of qubits. The purity of the reduced state of qubits in $A$ or $B$ is minimised when the number of sets of stabilizer generators is given by
\begin{equation}
k=2^{\min(|A|,|B|)}.
\end{equation}
This gives a new set of stabilizer generators for each measurement outcome when tracing out qubits, up to the number of measurement outcomes possible for the smaller set of $A$ and $B$ due to Eq.\ (\ref{SD}). Therefore the maximum value that the Concentratable Entanglement can achieve for a graph state of $n$ qubits is
\begin{equation}
\label{maxCE}
C_{\ket{G}}(\mathcal{S}) = 1 - \frac{1}{2^n}\sum_{j=0}^{n} {n \choose j} \frac{1}{2^{\min(j,n-j)}}.
\end{equation}

\begin{figure}[t!]
\begin{center}
\includegraphics[width=0.35\textwidth]{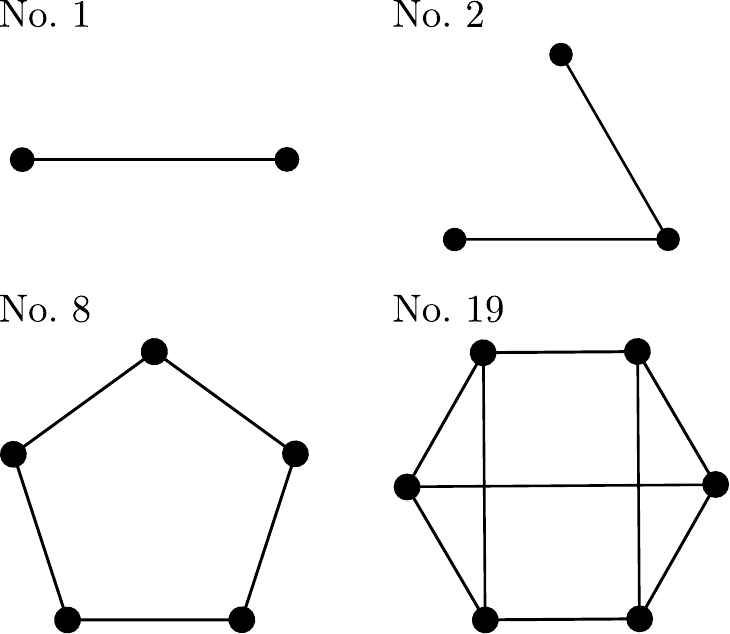}
\caption{The LU classes of graph states, numbered according to \cite{heinMultipartyEntanglementGraph2004, danielsenClassificationAllSelfDual2006, cabelloOptimalPreparationGraph2011}, where the overall Concentratable Entanglement achieves the maximum theoretical value.}
\label{Maximum}
\end{center}
\end{figure}
Graph states that maximise the Concentratable Entanglement exist only for 2, 3, 5 and 6 qubits, and an example from each LU equivalence class is shown in Fig.\ \ref{Maximum}. Trivially, a graph state of a single qubit satisfies Eq.\ (\ref{maxCE}), however it does not represent an entangled state and subsequently is of little interest. For graph states of 2 or 3 qubits, the maximum value of Concentratable Entanglement is equal to the minimum value, as the smaller (non-empty) bipartition always contains a single qubit, restricting all purities to the same value of $1/2$.

An absolutely maximally entangled (AME) state \cite{helwigAbsoluteMaximalEntanglement2012} has maximal entanglement in all bipartitions. If $B$ is the smaller set containing $j$ qubits in a bipartition $(A,B)$, then the bipartition is maximally entangled if the reduced state of qubits in $B$ can be written as the mixed state
\begin{equation}
\rho_B = \frac{1}{2^j}I_{2^j}.
\end{equation}
The purity of this state is $2^{-j}$. An AME state will achieve the maximum value of Concentratable Entanglement. The states of 2, 3, 5 and 6 qubits that maximise the Concentratable Entanglement are known AME states.

\begin{figure}[t!]
\begin{center}
\includegraphics[width=0.49\textwidth]{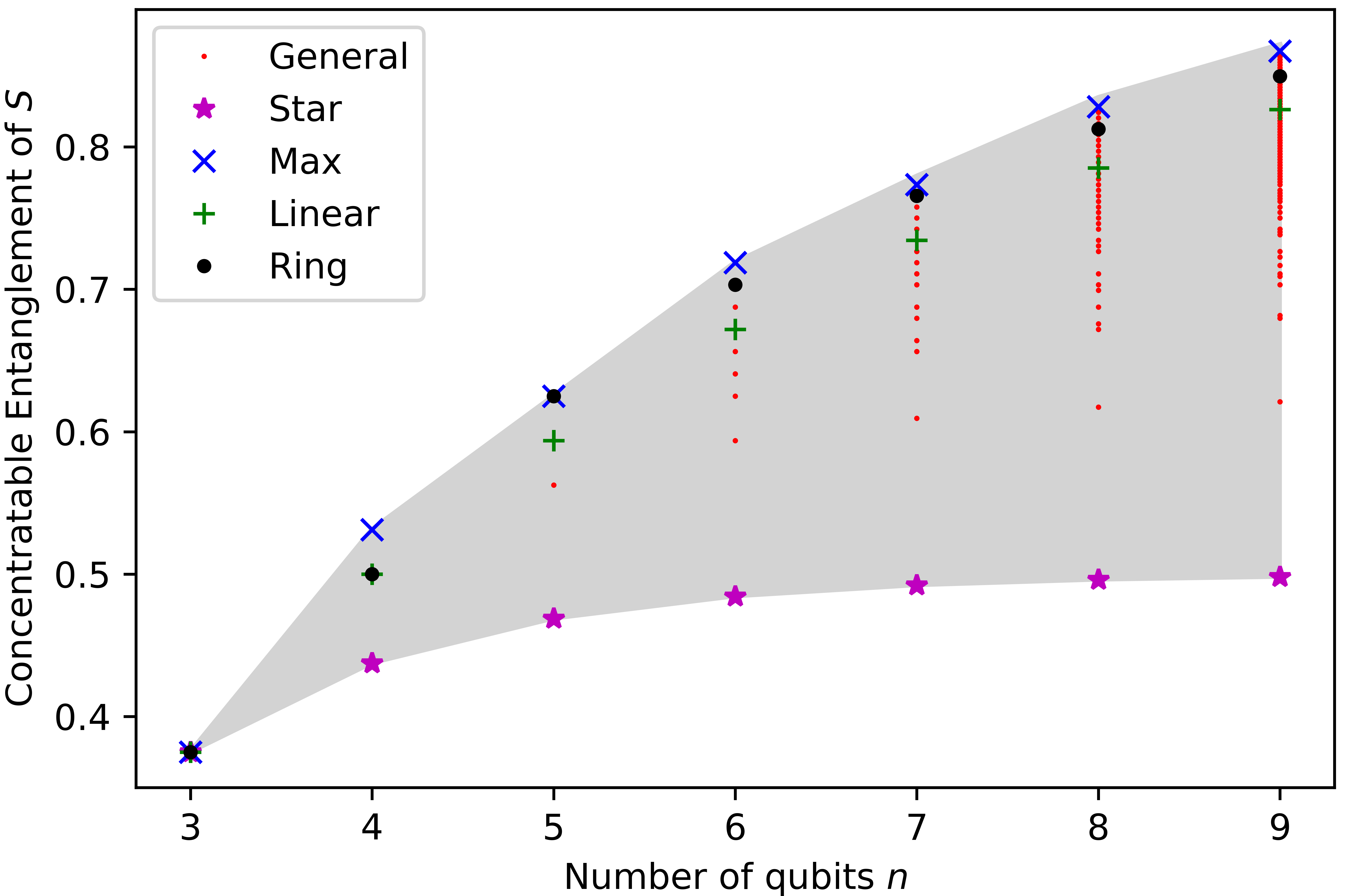}
\caption{The possible values of overall Concentratable Entanglement of graph states of 3 to 9 qubits. Theoretical bounds on the Concentratable Entanglement define the area in grey. We consider special classes of states, which are given different markers, as shown in the legend.}
\label{3-9}
\end{center}
\end{figure}

\textbf{5. Graphs up to nine qubits.}
For graph states consisting of nine qubits or fewer, we have calculated the overall Concentratable Entanglement of the state, as shown in Fig.\ \ref{3-9}. The respective values of 0 and 1/4 for the Concentratable Entanglement of the unique graph states of 1 and 2 qubits are not shown. The theoretical maximum and minimum values of Concentratable Entanglement bound the area shaded in grey. Star states achieving the minimum Concentratable Entanglement, which are LU equivalent to GHZ states lie on the lower boundary of this area. The states with the greatest Concentratable Entanglement of each size are shown, and can be seen to achieve the maximum only for states of 3, 5 and 6 qubits.

Figure \ref{3-9} also highlights the Concentratable Entanglement for linear and ring graph states. Linear cluster states of photons have been experimentally realised \cite{schwartzDeterministicGenerationCluster2016, istratiSequentialGenerationLinear2020}: an important step in developing the two dimensional clusters needed for MBQC and therefore their overall Concentratable Entanglement is shown. Ring states are similar to linear states although a further CZ operation is applied between the first and final qubits in the chain, thus completing a ring. Ring states have higher Concentratable Entanglement than linear states except for states of up to four qubits, where a ring and linear state are LU equivalent.

All of the overall Concentratable Entanglement values for graph states of nine qubits or fewer are shown in Fig.\ \ref{3-9}. Graph states within the same LU class will have the same overall Concentratable Entanglement, however using Concentratable Entanglement as a measure of the overall entanglement in a state is not sufficient to identify the LU class of a graph state, as the same value can occur for more than one class. For example, for graph states of seven qubits, there are 26 LU equivalence classes, but only 16 possible values of Concentratable Entanglement. If Concentratable Entanglement is calculated using Eq.\ (\ref{CE}), the reduced state purities can be studied to identify the LU class of a state. However, if a quantum computer were used to perform a SWAP test, further calculations would be required, such as performing SWAP tests to calculate the Concentratable Entanglement for subsets of qubits within the state to identify the LU class from the entanglement structure.

\section{Discussion \& Further Work}
We have presented a method for calculating the purity of reduced states and hence the Concentratable Entanglement for graph states entirely within the stabilizer formalism, using only the stabilizer generators of a graph state. By evaluating the Concentratable Entanglement for different graph states we have found closed forms when considering a single qubit in any graph state and either the inner or outer leaf qubits in a snowflake state for use in a quantum repeater. We have also shown how Concentratable Entanglement relates to previous results regarding Schmidt rank and AME states. However the overall Concentratable Entanglement of graph states in different LU classes is shown to overlap in certain cases, suggesting that calculation of this version of the measure alone is insufficient to identify the entanglement structure of a graph state.

Though Concentratable Entanglement was introduced for pure qubit states, it may be possible to use it as an entanglement measure for pure qudit states. Further work could then look to extend the definition in terms of purities given in Eq.\ (\ref{CE}) to graph states of qudits. Extension of the SWAP test to qudits \cite{proveExtendingControlled2022} may allow for calculation of Concentratable Entanglement via test outcomes when using a quantum computer in a similar manner to the qubit case. Qudit graph states \cite{looiQuantumErrorCorrecting2008, keetQuantumSecretSharing2010, helwigAbsolutelyMaximallyEntangled2013} are stabilizer states for qudits of prime dimension $d$ and allow for weighted or multiple edges. Reduced states could be found by considering all possible measurement outcomes, setting the stabilizer generators for measured qubits accordingly, and using the group structure of the stabilizer. The purity of the reduced state could again be calculated based on the number of unique sets of stabilizer generators.

It is unlikely that the result for calculating purities using stabilizer generators within the stabilizer formalism can be applied to hypergraphs.  Like qudit graph states, hypergraphs \cite{rossiQuantumHypergraphStates2013, quEncodingHypergraphsQuantum2013, steinhoffQuditHypergraphStates2017} allow for weighted or repeated edges between the same qudits but hyperedges that can connect any number of qudits are introduced. The stabilizer generator for each qudit is given in terms of CZ gates on multiple qudits, which are not within the Pauli group. Therefore hypergraphs with hyperedges connecting more than two qudits are not true stabilizer states and do not allow for the group properties of a stabilizer to be used to find updated generators following a measurement.

Currently Concentratable Entanglement is a measure that applies only to pure states. Even single qubit errors either in the generation or transmission of qubits within graph states can result in a mixed state. An extension of Concentratable Entanglement to mixed states could enable the study of effects such errors have on the entanglement within graph states. This is the subject of future investigations.

\medskip

\section{Acknowledgements}
ARC is supported by an EPSRC Doctoral Training Partnership (DTP) Scholarship.
PK is supported by the EPSRC Quantum Communications Hub, Grant No.\ EP/M013472/1, and the EPSRC grant Large Baseline Quantum-Enhanced Imaging Networks, Grant No.\
EP/V021303/1.


\begin{thebibliography}{39}%
\makeatletter
\providecommand \@ifxundefined [1]{%
 \@ifx{#1\undefined}
}%
\providecommand \@ifnum [1]{%
 \ifnum #1\expandafter \@firstoftwo
 \else \expandafter \@secondoftwo
 \fi
}%
\providecommand \@ifx [1]{%
 \ifx #1\expandafter \@firstoftwo
 \else \expandafter \@secondoftwo
 \fi
}%
\providecommand \natexlab [1]{#1}%
\providecommand \enquote  [1]{``#1''}%
\providecommand \bibnamefont  [1]{#1}%
\providecommand \bibfnamefont [1]{#1}%
\providecommand \citenamefont [1]{#1}%
\providecommand \href@noop [0]{\@secondoftwo}%
\providecommand \href [0]{\begingroup \@sanitize@url \@href}%
\providecommand \@href[1]{\@@startlink{#1}\@@href}%
\providecommand \@@href[1]{\endgroup#1\@@endlink}%
\providecommand \@sanitize@url [0]{\catcode `\\12\catcode `\$12\catcode
  `\&12\catcode `\#12\catcode `\^12\catcode `\_12\catcode `\%12\relax}%
\providecommand \@@startlink[1]{}%
\providecommand \@@endlink[0]{}%
\providecommand \url  [0]{\begingroup\@sanitize@url \@url }%
\providecommand \@url [1]{\endgroup\@href {#1}{\urlprefix }}%
\providecommand \urlprefix  [0]{URL }%
\providecommand \Eprint [0]{\href }%
\providecommand \doibase [0]{http://dx.doi.org/}%
\providecommand \selectlanguage [0]{\@gobble}%
\providecommand \bibinfo  [0]{\@secondoftwo}%
\providecommand \bibfield  [0]{\@secondoftwo}%
\providecommand \translation [1]{[#1]}%
\providecommand \BibitemOpen [0]{}%
\providecommand \bibitemStop [0]{}%
\providecommand \bibitemNoStop [0]{.\EOS\space}%
\providecommand \EOS [0]{\spacefactor3000\relax}%
\providecommand \BibitemShut  [1]{\csname bibitem#1\endcsname}%
\let\auto@bib@innerbib\@empty
%</preamble>
\bibitem [{\citenamefont {Einstein}\ \emph {et~al.}(1935)\citenamefont
  {Einstein}, \citenamefont {Podolsky},\ and\ \citenamefont
  {Rosen}}]{einsteinCanQuantumMechanicalDescription1935}%
  \BibitemOpen
  \bibfield  {author} {\bibinfo {author} {\bibfnamefont {A.}~\bibnamefont
  {Einstein}}, \bibinfo {author} {\bibfnamefont {B.}~\bibnamefont {Podolsky}},
  \ and\ \bibinfo {author} {\bibfnamefont {N.}~\bibnamefont {Rosen}},\ }\href
  {\doibase 10.1103/PhysRev.47.777} {\bibfield  {journal} {\bibinfo  {journal}
  {Phys. Rev.}\ }\textbf {\bibinfo {volume} {47}},\ \bibinfo {pages} {777}
  (\bibinfo {year} {1935})}\BibitemShut {NoStop}%
\bibitem [{\citenamefont
  {Schr{\"o}dinger}(1935)}]{schrodingerDiscussionProbabilityRelations1935}%
  \BibitemOpen
  \bibfield  {author} {\bibinfo {author} {\bibfnamefont {E.}~\bibnamefont
  {Schr{\"o}dinger}},\ }\href {\doibase 10.1017/S0305004100013554} {\bibfield
  {journal} {\bibinfo  {journal} {Math. Proc. Camb. Phil. Soc.}\ }\textbf
  {\bibinfo {volume} {31}},\ \bibinfo {pages} {555} (\bibinfo {year}
  {1935})}\BibitemShut {NoStop}%
\bibitem [{\citenamefont {Hein}\ \emph {et~al.}(2004)\citenamefont {Hein},
  \citenamefont {Eisert},\ and\ \citenamefont
  {Briegel}}]{heinMultipartyEntanglementGraph2004}%
  \BibitemOpen
  \bibfield  {author} {\bibinfo {author} {\bibfnamefont {M.}~\bibnamefont
  {Hein}}, \bibinfo {author} {\bibfnamefont {J.}~\bibnamefont {Eisert}}, \ and\
  \bibinfo {author} {\bibfnamefont {H.~J.}\ \bibnamefont {Briegel}},\ }\href
  {\doibase 10.1103/PhysRevA.69.062311} {\bibfield  {journal} {\bibinfo
  {journal} {Phys. Rev. A}\ }\textbf {\bibinfo {volume} {69}},\ \bibinfo
  {pages} {062311} (\bibinfo {year} {2004})}\BibitemShut {NoStop}%
\bibitem [{\citenamefont {Briegel}\ \emph {et~al.}(2009)\citenamefont
  {Briegel}, \citenamefont {Browne}, \citenamefont {Dür}, \citenamefont
  {Raussendorf},\ and\ \citenamefont {Nest}}]{briegelMeasurementBased2009}%
  \BibitemOpen
  \bibfield  {author} {\bibinfo {author} {\bibfnamefont {H.~J.}\ \bibnamefont
  {Briegel}}, \bibinfo {author} {\bibfnamefont {D.~E.}\ \bibnamefont {Browne}},
  \bibinfo {author} {\bibfnamefont {W.}~\bibnamefont {Dür}}, \bibinfo {author}
  {\bibfnamefont {R.}~\bibnamefont {Raussendorf}}, \ and\ \bibinfo {author}
  {\bibfnamefont {M.~V.~d.}\ \bibnamefont {Nest}},\ }\href@noop {} {\bibfield
  {journal} {\bibinfo  {journal} {Nat. Phys.}\ }\textbf {\bibinfo {volume}
  {5}},\ \bibinfo {pages} {8} (\bibinfo {year} {2009})}\BibitemShut {NoStop}%
\bibitem [{\citenamefont {Raussendorf}\ and\ \citenamefont
  {Briegel}(2001)}]{raussendorfOneWayQuantumComputer2001}%
  \BibitemOpen
  \bibfield  {author} {\bibinfo {author} {\bibfnamefont {R.}~\bibnamefont
  {Raussendorf}}\ and\ \bibinfo {author} {\bibfnamefont {H.~J.}\ \bibnamefont
  {Briegel}},\ }\href {\doibase 10.1103/PhysRevLett.86.5188} {\bibfield
  {journal} {\bibinfo  {journal} {Phys. Rev. Lett.}\ }\textbf {\bibinfo
  {volume} {86}},\ \bibinfo {pages} {5188} (\bibinfo {year}
  {2001})}\BibitemShut {NoStop}%
\bibitem [{\citenamefont {Raussendorf}\ \emph {et~al.}(2003)\citenamefont
  {Raussendorf}, \citenamefont {Browne},\ and\ \citenamefont
  {Briegel}}]{raussendorfMeasurementbasedQuantumComputation2003}%
  \BibitemOpen
  \bibfield  {author} {\bibinfo {author} {\bibfnamefont {R.}~\bibnamefont
  {Raussendorf}}, \bibinfo {author} {\bibfnamefont {D.~E.}\ \bibnamefont
  {Browne}}, \ and\ \bibinfo {author} {\bibfnamefont {H.~J.}\ \bibnamefont
  {Briegel}},\ }\href {\doibase 10.1103/PhysRevA.68.022312} {\bibfield
  {journal} {\bibinfo  {journal} {Phys. Rev. A}\ }\textbf {\bibinfo {volume}
  {68}},\ \bibinfo {pages} {022312} (\bibinfo {year} {2003})}\BibitemShut
  {NoStop}%
\bibitem [{\citenamefont {Briegel}\ \emph {et~al.}(1998)\citenamefont
  {Briegel}, \citenamefont {D{\"u}r}, \citenamefont {Cirac},\ and\
  \citenamefont {Zoller}}]{briegelQuantumRepeatersRole1998}%
  \BibitemOpen
  \bibfield  {author} {\bibinfo {author} {\bibfnamefont {H.-J.}\ \bibnamefont
  {Briegel}}, \bibinfo {author} {\bibfnamefont {W.}~\bibnamefont {D{\"u}r}},
  \bibinfo {author} {\bibfnamefont {J.~I.}\ \bibnamefont {Cirac}}, \ and\
  \bibinfo {author} {\bibfnamefont {P.}~\bibnamefont {Zoller}},\ }\href
  {\doibase 10.1103/PhysRevLett.81.5932} {\bibfield  {journal} {\bibinfo
  {journal} {Phys. Rev. Lett.}\ }\textbf {\bibinfo {volume} {81}},\ \bibinfo
  {pages} {5932} (\bibinfo {year} {1998})}\BibitemShut {NoStop}%
\bibitem [{\citenamefont {Zwerger}\ \emph {et~al.}(2012)\citenamefont
  {Zwerger}, \citenamefont {D{\"u}r},\ and\ \citenamefont
  {Briegel}}]{zwergerMeasurementbasedQuantumRepeaters2012}%
  \BibitemOpen
  \bibfield  {author} {\bibinfo {author} {\bibfnamefont {M.}~\bibnamefont
  {Zwerger}}, \bibinfo {author} {\bibfnamefont {W.}~\bibnamefont {D{\"u}r}}, \
  and\ \bibinfo {author} {\bibfnamefont {H.~J.}\ \bibnamefont {Briegel}},\
  }\href {\doibase 10.1103/PhysRevA.85.062326} {\bibfield  {journal} {\bibinfo
  {journal} {Phys. Rev. A}\ }\textbf {\bibinfo {volume} {85}},\ \bibinfo
  {pages} {062326} (\bibinfo {year} {2012})}\BibitemShut {NoStop}%
\bibitem [{\citenamefont {Azuma}\ \emph {et~al.}(2015)\citenamefont {Azuma},
  \citenamefont {Tamaki},\ and\ \citenamefont
  {Lo}}]{azumaAllphotonicQuantumRepeaters2015}%
  \BibitemOpen
  \bibfield  {author} {\bibinfo {author} {\bibfnamefont {K.}~\bibnamefont
  {Azuma}}, \bibinfo {author} {\bibfnamefont {K.}~\bibnamefont {Tamaki}}, \
  and\ \bibinfo {author} {\bibfnamefont {H.-K.}\ \bibnamefont {Lo}},\ }\href
  {\doibase 10.1038/ncomms7787} {\bibfield  {journal} {\bibinfo  {journal}
  {Nat. Commun.}\ }\textbf {\bibinfo {volume} {6}},\ \bibinfo {pages} {6787}
  (\bibinfo {year} {2015})}\BibitemShut {NoStop}%
\bibitem [{\citenamefont
  {Gottesman}(1997)}]{gottesmanStabilizerCodesQuantum1997}%
  \BibitemOpen
  \bibfield  {author} {\bibinfo {author} {\bibfnamefont {D.}~\bibnamefont
  {Gottesman}},\ }\href@noop {} {\enquote {\bibinfo {title} {Stabilizer
  {{Codes}} and {{Quantum Error Correction}}},}\ } (\bibinfo {year} {1997}),\
  \bibinfo {note} {{Caltech Ph.D. Thesis}},\ \Eprint
  {http://arxiv.org/abs/quant-ph/9705052} {arXiv:quant-ph/9705052} \BibitemShut
  {NoStop}%
\bibitem [{\citenamefont {Schlingemann}\ and\ \citenamefont
  {Werner}(2001)}]{schlingemannQuantumErrorCorrecting2001}%
  \BibitemOpen
  \bibfield  {author} {\bibinfo {author} {\bibfnamefont {D.}~\bibnamefont
  {Schlingemann}}\ and\ \bibinfo {author} {\bibfnamefont {R.~F.}\ \bibnamefont
  {Werner}},\ }\href {\doibase 10.1103/PhysRevA.65.012308} {\bibfield
  {journal} {\bibinfo  {journal} {Phys. Rev. A}\ }\textbf {\bibinfo {volume}
  {65}},\ \bibinfo {pages} {012308} (\bibinfo {year} {2001})}\BibitemShut
  {NoStop}%
\bibitem [{\citenamefont {Markham}\ and\ \citenamefont
  {Sanders}(2008)}]{markhamGraphStatesQuantum2008}%
  \BibitemOpen
  \bibfield  {author} {\bibinfo {author} {\bibfnamefont {D.}~\bibnamefont
  {Markham}}\ and\ \bibinfo {author} {\bibfnamefont {B.~C.}\ \bibnamefont
  {Sanders}},\ }\href {\doibase 10.1103/PhysRevA.78.042309} {\bibfield
  {journal} {\bibinfo  {journal} {Phys. Rev. A}\ }\textbf {\bibinfo {volume}
  {78}},\ \bibinfo {pages} {042309} (\bibinfo {year} {2008})}\BibitemShut
  {NoStop}%
\bibitem [{\citenamefont
  {Peres}(1996)}]{peresSeparabilityCriterionDensity1996}%
  \BibitemOpen
  \bibfield  {author} {\bibinfo {author} {\bibfnamefont {A.}~\bibnamefont
  {Peres}},\ }\href {\doibase 10.1103/PhysRevLett.77.1413} {\bibfield
  {journal} {\bibinfo  {journal} {Phys. Rev. Lett.}\ }\textbf {\bibinfo
  {volume} {77}},\ \bibinfo {pages} {1413} (\bibinfo {year}
  {1996})}\BibitemShut {NoStop}%
\bibitem [{\citenamefont {Horodecki}\ \emph {et~al.}(1996)\citenamefont
  {Horodecki}, \citenamefont {Horodecki},\ and\ \citenamefont
  {Horodecki}}]{horodeckiSeparabilityMixedStates1996}%
  \BibitemOpen
  \bibfield  {author} {\bibinfo {author} {\bibfnamefont {M.}~\bibnamefont
  {Horodecki}}, \bibinfo {author} {\bibfnamefont {P.}~\bibnamefont
  {Horodecki}}, \ and\ \bibinfo {author} {\bibfnamefont {R.}~\bibnamefont
  {Horodecki}},\ }\href {\doibase 10.1016/S0375-9601(96)00706-2} {\bibfield
  {journal} {\bibinfo  {journal} {Physics Letters A}\ }\textbf {\bibinfo
  {volume} {223}},\ \bibinfo {pages} {1} (\bibinfo {year} {1996})},\ \bibinfo
  {note} {arXiv:quant-ph/9605038}\BibitemShut {NoStop}%
\bibitem [{\citenamefont {Bennett}\ \emph {et~al.}(1996)\citenamefont
  {Bennett}, \citenamefont {DiVincenzo}, \citenamefont {Smolin},\ and\
  \citenamefont {Wootters}}]{bennettMixedstateEntanglementQuantum1996}%
  \BibitemOpen
  \bibfield  {author} {\bibinfo {author} {\bibfnamefont {C.~H.}\ \bibnamefont
  {Bennett}}, \bibinfo {author} {\bibfnamefont {D.~P.}\ \bibnamefont
  {DiVincenzo}}, \bibinfo {author} {\bibfnamefont {J.~A.}\ \bibnamefont
  {Smolin}}, \ and\ \bibinfo {author} {\bibfnamefont {W.~K.}\ \bibnamefont
  {Wootters}},\ }\href {\doibase 10.1103/PhysRevA.54.3824} {\bibfield
  {journal} {\bibinfo  {journal} {Phys. Rev. A}\ }\textbf {\bibinfo {volume}
  {54}},\ \bibinfo {pages} {3824} (\bibinfo {year} {1996})}\BibitemShut
  {NoStop}%
\bibitem [{\citenamefont {Hill}\ and\ \citenamefont
  {Wootters}(1997)}]{hillEntanglementPairQuantum1997}%
  \BibitemOpen
  \bibfield  {author} {\bibinfo {author} {\bibfnamefont {S.}~\bibnamefont
  {Hill}}\ and\ \bibinfo {author} {\bibfnamefont {W.~K.}\ \bibnamefont
  {Wootters}},\ }\href {\doibase 10.1103/PhysRevLett.78.5022} {\bibfield
  {journal} {\bibinfo  {journal} {Phys. Rev. Lett.}\ }\textbf {\bibinfo
  {volume} {78}},\ \bibinfo {pages} {5022} (\bibinfo {year}
  {1997})}\BibitemShut {NoStop}%
\bibitem [{\citenamefont {Rungta}\ \emph {et~al.}(2001)\citenamefont {Rungta},
  \citenamefont {Bu{\v z}ek}, \citenamefont {Caves}, \citenamefont {Hillery},\
  and\ \citenamefont {Milburn}}]{rungtaUniversalStateInversion2001}%
  \BibitemOpen
  \bibfield  {author} {\bibinfo {author} {\bibfnamefont {P.}~\bibnamefont
  {Rungta}}, \bibinfo {author} {\bibfnamefont {V.}~\bibnamefont {Bu{\v z}ek}},
  \bibinfo {author} {\bibfnamefont {C.~M.}\ \bibnamefont {Caves}}, \bibinfo
  {author} {\bibfnamefont {M.}~\bibnamefont {Hillery}}, \ and\ \bibinfo
  {author} {\bibfnamefont {G.~J.}\ \bibnamefont {Milburn}},\ }\href {\doibase
  10.1103/PhysRevA.64.042315} {\bibfield  {journal} {\bibinfo  {journal} {Phys.
  Rev. A}\ }\textbf {\bibinfo {volume} {64}},\ \bibinfo {pages} {042315}
  (\bibinfo {year} {2001})}\BibitemShut {NoStop}%
\bibitem [{\citenamefont {D{\"u}r}\ \emph {et~al.}(2000)\citenamefont
  {D{\"u}r}, \citenamefont {Vidal},\ and\ \citenamefont
  {Cirac}}]{durThreeQubitsCan2000}%
  \BibitemOpen
  \bibfield  {author} {\bibinfo {author} {\bibfnamefont {W.}~\bibnamefont
  {D{\"u}r}}, \bibinfo {author} {\bibfnamefont {G.}~\bibnamefont {Vidal}}, \
  and\ \bibinfo {author} {\bibfnamefont {J.~I.}\ \bibnamefont {Cirac}},\ }\href
  {\doibase 10.1103/PhysRevA.62.062314} {\bibfield  {journal} {\bibinfo
  {journal} {Phys. Rev. A}\ }\textbf {\bibinfo {volume} {62}},\ \bibinfo
  {pages} {062314} (\bibinfo {year} {2000})}\BibitemShut {NoStop}%
\bibitem [{\citenamefont {Meyer}\ and\ \citenamefont
  {Wallach}(2002)}]{meyerGlobalEntanglementMultiparticle2002}%
  \BibitemOpen
  \bibfield  {author} {\bibinfo {author} {\bibfnamefont {D.~A.}\ \bibnamefont
  {Meyer}}\ and\ \bibinfo {author} {\bibfnamefont {N.~R.}\ \bibnamefont
  {Wallach}},\ }\href {\doibase 10.1063/1.1497700} {\bibfield  {journal}
  {\bibinfo  {journal} {J. Math. Phys.}\ }\textbf {\bibinfo {volume} {43}},\
  \bibinfo {pages} {4273} (\bibinfo {year} {2002})}\BibitemShut {NoStop}%
\bibitem [{\citenamefont
  {Brennen}(2003)}]{brennenObservableMeasureEntanglement2003}%
  \BibitemOpen
  \bibfield  {author} {\bibinfo {author} {\bibfnamefont {G.}~\bibnamefont
  {Brennen}},\ }\href {\doibase 10.26421/QIC3.6-5} {\bibfield  {journal}
  {\bibinfo  {journal} {Quantum Inf. Comput.}\ }\textbf {\bibinfo {volume}
  {3}},\ \bibinfo {pages} {619} (\bibinfo {year} {2003})}\BibitemShut {NoStop}%
\bibitem [{\citenamefont {Carvalho}\ \emph {et~al.}(2004)\citenamefont
  {Carvalho}, \citenamefont {Mintert},\ and\ \citenamefont
  {Buchleitner}}]{carvalhoDecoherenceMultipartiteEntanglement2004}%
  \BibitemOpen
  \bibfield  {author} {\bibinfo {author} {\bibfnamefont {A.~R.~R.}\
  \bibnamefont {Carvalho}}, \bibinfo {author} {\bibfnamefont {F.}~\bibnamefont
  {Mintert}}, \ and\ \bibinfo {author} {\bibfnamefont {A.}~\bibnamefont
  {Buchleitner}},\ }\href {\doibase 10.1103/PhysRevLett.93.230501} {\bibfield
  {journal} {\bibinfo  {journal} {Phys. Rev. Lett.}\ }\textbf {\bibinfo
  {volume} {93}},\ \bibinfo {pages} {230501} (\bibinfo {year}
  {2004})}\BibitemShut {NoStop}%
\bibitem [{\citenamefont {Beckey}\ \emph {et~al.}(2021)\citenamefont {Beckey},
  \citenamefont {Gigena}, \citenamefont {Coles},\ and\ \citenamefont
  {Cerezo}}]{beckeyComputableOperationallyMeaningful2021}%
  \BibitemOpen
  \bibfield  {author} {\bibinfo {author} {\bibfnamefont {J.~L.}\ \bibnamefont
  {Beckey}}, \bibinfo {author} {\bibfnamefont {N.}~\bibnamefont {Gigena}},
  \bibinfo {author} {\bibfnamefont {P.~J.}\ \bibnamefont {Coles}}, \ and\
  \bibinfo {author} {\bibfnamefont {M.}~\bibnamefont {Cerezo}},\ }\href
  {\doibase 10.1103/PhysRevLett.127.140501} {\bibfield  {journal} {\bibinfo
  {journal} {Phys. Rev. Lett.}\ }\textbf {\bibinfo {volume} {127}},\ \bibinfo
  {pages} {140501} (\bibinfo {year} {2021})}\BibitemShut {NoStop}%
\bibitem [{\citenamefont {Foulds}\ \emph {et~al.}(2021)\citenamefont {Foulds},
  \citenamefont {Kendon},\ and\ \citenamefont
  {Spiller}}]{fouldsControlledSWAPTest2021}%
  \BibitemOpen
  \bibfield  {author} {\bibinfo {author} {\bibfnamefont {S.}~\bibnamefont
  {Foulds}}, \bibinfo {author} {\bibfnamefont {V.}~\bibnamefont {Kendon}}, \
  and\ \bibinfo {author} {\bibfnamefont {T.}~\bibnamefont {Spiller}},\ }\href
  {\doibase 10.1088/2058-9565/abe458} {\bibfield  {journal} {\bibinfo
  {journal} {Quantum Sci. Technol.}\ }\textbf {\bibinfo {volume} {6}},\
  \bibinfo {pages} {035002} (\bibinfo {year} {2021})}\BibitemShut {NoStop}%
\bibitem [{\citenamefont {Wong}\ and\ \citenamefont
  {Christensen}(2001)}]{wongPotentialMultiparticleEntanglement2001}%
  \BibitemOpen
  \bibfield  {author} {\bibinfo {author} {\bibfnamefont {A.}~\bibnamefont
  {Wong}}\ and\ \bibinfo {author} {\bibfnamefont {N.}~\bibnamefont
  {Christensen}},\ }\href {\doibase 10.1103/PhysRevA.63.044301} {\bibfield
  {journal} {\bibinfo  {journal} {Phys. Rev. A}\ }\textbf {\bibinfo {volume}
  {63}},\ \bibinfo {pages} {044301} (\bibinfo {year} {2001})}\BibitemShut
  {NoStop}%
\bibitem [{\citenamefont {Hein}\ \emph {et~al.}(2006)\citenamefont {Hein},
  \citenamefont {D{\"u}r}, \citenamefont {Eisert}, \citenamefont {Raussendorf},
  \citenamefont {den Nest},\ and\ \citenamefont
  {Briegel}}]{heinEntanglementGraphStates2006}%
  \BibitemOpen
  \bibfield  {author} {\bibinfo {author} {\bibfnamefont {M.}~\bibnamefont
  {Hein}}, \bibinfo {author} {\bibfnamefont {W.}~\bibnamefont {D{\"u}r}},
  \bibinfo {author} {\bibfnamefont {J.}~\bibnamefont {Eisert}}, \bibinfo
  {author} {\bibfnamefont {R.}~\bibnamefont {Raussendorf}}, \bibinfo {author}
  {\bibfnamefont {M.~V.}\ \bibnamefont {den Nest}}, \ and\ \bibinfo {author}
  {\bibfnamefont {H.-J.}\ \bibnamefont {Briegel}},\ }in\ \href@noop {} {\emph
  {\bibinfo {booktitle} {Proceedings of the {International} {School} of
  {Physics} {``E. Fermi"}}}}\ (\bibinfo {year} {2006})\ pp.\ \bibinfo {pages}
  {115--218}\BibitemShut {NoStop}%
\bibitem [{\citenamefont {Buhrman}\ \emph {et~al.}(2001)\citenamefont
  {Buhrman}, \citenamefont {Cleve}, \citenamefont {Watrous},\ and\
  \citenamefont {{de Wolf}}}]{buhrmanQuantumFingerprinting2001}%
  \BibitemOpen
  \bibfield  {author} {\bibinfo {author} {\bibfnamefont {H.}~\bibnamefont
  {Buhrman}}, \bibinfo {author} {\bibfnamefont {R.}~\bibnamefont {Cleve}},
  \bibinfo {author} {\bibfnamefont {J.}~\bibnamefont {Watrous}}, \ and\
  \bibinfo {author} {\bibfnamefont {R.}~\bibnamefont {{de Wolf}}},\ }\href
  {\doibase 10.1103/PhysRevLett.87.167902} {\bibfield  {journal} {\bibinfo
  {journal} {Phys. Rev. Lett.}\ }\textbf {\bibinfo {volume} {87}},\ \bibinfo
  {pages} {167902} (\bibinfo {year} {2001})}\BibitemShut {NoStop}%
\bibitem [{\citenamefont {Nielsen}\ and\ \citenamefont
  {Chuang}(2010)}]{nielsenQuantumComputationQuantum2010}%
  \BibitemOpen
  \bibfield  {author} {\bibinfo {author} {\bibfnamefont {M.}~\bibnamefont
  {Nielsen}}\ and\ \bibinfo {author} {\bibfnamefont {I.}~\bibnamefont
  {Chuang}},\ }\href@noop {} {\emph {\bibinfo {title} {Quantum {{Computation}}
  and {{Quantum Information}}}}},\ \bibinfo {edition} {{10th Anniversary}}\
  ed.\ (\bibinfo  {publisher} {{Cambridge University Press}},\ \bibinfo {year}
  {2010})\BibitemShut {NoStop}%
\bibitem [{\citenamefont {Danielsen}\ and\ \citenamefont
  {Parker}(2006)}]{danielsenClassificationAllSelfDual2006}%
  \BibitemOpen
  \bibfield  {author} {\bibinfo {author} {\bibfnamefont {L.~E.}\ \bibnamefont
  {Danielsen}}\ and\ \bibinfo {author} {\bibfnamefont {M.~G.}\ \bibnamefont
  {Parker}},\ }\href {\doibase 10.1016/j.jcta.2005.12.004} {\bibfield
  {journal} {\bibinfo  {journal} {J. of Comb. Theory, Series A}\ }\textbf
  {\bibinfo {volume} {113}},\ \bibinfo {pages} {1351} (\bibinfo {year}
  {2006})}\BibitemShut {NoStop}%
\bibitem [{\citenamefont {Cabello}\ \emph {et~al.}(2011)\citenamefont
  {Cabello}, \citenamefont {Danielsen}, \citenamefont {{L{\'o}pez-Tarrida}},\
  and\ \citenamefont {Portillo}}]{cabelloOptimalPreparationGraph2011}%
  \BibitemOpen
  \bibfield  {author} {\bibinfo {author} {\bibfnamefont {A.}~\bibnamefont
  {Cabello}}, \bibinfo {author} {\bibfnamefont {L.~E.}\ \bibnamefont
  {Danielsen}}, \bibinfo {author} {\bibfnamefont {A.~J.}\ \bibnamefont
  {{L{\'o}pez-Tarrida}}}, \ and\ \bibinfo {author} {\bibfnamefont {J.~R.}\
  \bibnamefont {Portillo}},\ }\href {\doibase 10.1103/PhysRevA.83.042314}
  {\bibfield  {journal} {\bibinfo  {journal} {Phys. Rev. A}\ }\textbf {\bibinfo
  {volume} {83}},\ \bibinfo {pages} {042314} (\bibinfo {year}
  {2011})}\BibitemShut {NoStop}%
\bibitem [{\citenamefont {Helwig}\ \emph {et~al.}(2012)\citenamefont {Helwig},
  \citenamefont {Cui}, \citenamefont {Riera}, \citenamefont {Latorre},\ and\
  \citenamefont {Lo}}]{helwigAbsoluteMaximalEntanglement2012}%
  \BibitemOpen
  \bibfield  {author} {\bibinfo {author} {\bibfnamefont {W.}~\bibnamefont
  {Helwig}}, \bibinfo {author} {\bibfnamefont {W.}~\bibnamefont {Cui}},
  \bibinfo {author} {\bibfnamefont {A.}~\bibnamefont {Riera}}, \bibinfo
  {author} {\bibfnamefont {J.~I.}\ \bibnamefont {Latorre}}, \ and\ \bibinfo
  {author} {\bibfnamefont {H.-K.}\ \bibnamefont {Lo}},\ }\href {\doibase
  10.1103/PhysRevA.86.052335} {\bibfield  {journal} {\bibinfo  {journal} {Phys.
  Rev. A}\ }\textbf {\bibinfo {volume} {86}},\ \bibinfo {pages} {052335}
  (\bibinfo {year} {2012})}\BibitemShut {NoStop}%
\bibitem [{\citenamefont {Schwartz}\ \emph {et~al.}(2016)\citenamefont
  {Schwartz}, \citenamefont {Cogan}, \citenamefont {Schmidgall}, \citenamefont
  {Don}, \citenamefont {Gantz}, \citenamefont {Kenneth}, \citenamefont
  {Lindner},\ and\ \citenamefont
  {Gershoni}}]{schwartzDeterministicGenerationCluster2016}%
  \BibitemOpen
  \bibfield  {author} {\bibinfo {author} {\bibfnamefont {I.}~\bibnamefont
  {Schwartz}}, \bibinfo {author} {\bibfnamefont {D.}~\bibnamefont {Cogan}},
  \bibinfo {author} {\bibfnamefont {E.~R.}\ \bibnamefont {Schmidgall}},
  \bibinfo {author} {\bibfnamefont {Y.}~\bibnamefont {Don}}, \bibinfo {author}
  {\bibfnamefont {L.}~\bibnamefont {Gantz}}, \bibinfo {author} {\bibfnamefont
  {O.}~\bibnamefont {Kenneth}}, \bibinfo {author} {\bibfnamefont {N.~H.}\
  \bibnamefont {Lindner}}, \ and\ \bibinfo {author} {\bibfnamefont
  {D.}~\bibnamefont {Gershoni}},\ }\href {\doibase 10.1126/science.aah4758}
  {\bibfield  {journal} {\bibinfo  {journal} {Science}\ }\textbf {\bibinfo
  {volume} {354}},\ \bibinfo {pages} {434} (\bibinfo {year}
  {2016})}\BibitemShut {NoStop}%
\bibitem [{\citenamefont {Istrati}\ \emph {et~al.}(2020)\citenamefont
  {Istrati}, \citenamefont {Pilnyak}, \citenamefont {Loredo}, \citenamefont
  {Antón}, \citenamefont {Somaschi}, \citenamefont {Hilaire}, \citenamefont
  {Ollivier}, \citenamefont {Esmann}, \citenamefont {Cohen}, \citenamefont
  {Vidro}, \citenamefont {Millet}, \citenamefont {Lemaître}, \citenamefont
  {Sagnes}, \citenamefont {Harouri}, \citenamefont {Lanco}, \citenamefont
  {Senellart},\ and\ \citenamefont
  {Eisenberg}}]{istratiSequentialGenerationLinear2020}%
  \BibitemOpen
  \bibfield  {author} {\bibinfo {author} {\bibfnamefont {D.}~\bibnamefont
  {Istrati}}, \bibinfo {author} {\bibfnamefont {Y.}~\bibnamefont {Pilnyak}},
  \bibinfo {author} {\bibfnamefont {J.~C.}\ \bibnamefont {Loredo}}, \bibinfo
  {author} {\bibfnamefont {C.}~\bibnamefont {Antón}}, \bibinfo {author}
  {\bibfnamefont {N.}~\bibnamefont {Somaschi}}, \bibinfo {author}
  {\bibfnamefont {P.}~\bibnamefont {Hilaire}}, \bibinfo {author} {\bibfnamefont
  {H.}~\bibnamefont {Ollivier}}, \bibinfo {author} {\bibfnamefont
  {M.}~\bibnamefont {Esmann}}, \bibinfo {author} {\bibfnamefont
  {L.}~\bibnamefont {Cohen}}, \bibinfo {author} {\bibfnamefont
  {L.}~\bibnamefont {Vidro}}, \bibinfo {author} {\bibfnamefont
  {C.}~\bibnamefont {Millet}}, \bibinfo {author} {\bibfnamefont
  {A.}~\bibnamefont {Lemaître}}, \bibinfo {author} {\bibfnamefont
  {I.}~\bibnamefont {Sagnes}}, \bibinfo {author} {\bibfnamefont
  {A.}~\bibnamefont {Harouri}}, \bibinfo {author} {\bibfnamefont
  {L.}~\bibnamefont {Lanco}}, \bibinfo {author} {\bibfnamefont
  {P.}~\bibnamefont {Senellart}}, \ and\ \bibinfo {author} {\bibfnamefont
  {H.~S.}\ \bibnamefont {Eisenberg}},\ }\href {\doibase
  10.1038/s41467-020-19341-4} {\bibfield  {journal} {\bibinfo  {journal} {Nat.
  Commun.}\ }\textbf {\bibinfo {volume} {11}},\ \bibinfo {pages} {5501}
  (\bibinfo {year} {2020})}\BibitemShut {NoStop}%
\bibitem [{\citenamefont {Prove}\ \emph {et~al.}(2022)\citenamefont {Prove},
  \citenamefont {Foulds},\ and\ \citenamefont
  {Kendon}}]{proveExtendingControlled2022}%
  \BibitemOpen
  \bibfield  {author} {\bibinfo {author} {\bibfnamefont {O.}~\bibnamefont
  {Prove}}, \bibinfo {author} {\bibfnamefont {S.}~\bibnamefont {Foulds}}, \
  and\ \bibinfo {author} {\bibfnamefont {V.}~\bibnamefont {Kendon}},\
  }\href@noop {} {\enquote {\bibinfo {title} {Extending the controlled {SWAP}
  test to higher dimensions},}\ } (\bibinfo {year} {2022}),\ \Eprint
  {http://arxiv.org/abs/2112.04333} {arXiv:2112.04333} \BibitemShut {NoStop}%
\bibitem [{\citenamefont {Looi}\ \emph {et~al.}(2008)\citenamefont {Looi},
  \citenamefont {Yu}, \citenamefont {Gheorghiu},\ and\ \citenamefont
  {Griffiths}}]{looiQuantumErrorCorrecting2008}%
  \BibitemOpen
  \bibfield  {author} {\bibinfo {author} {\bibfnamefont {S.~Y.}\ \bibnamefont
  {Looi}}, \bibinfo {author} {\bibfnamefont {L.}~\bibnamefont {Yu}}, \bibinfo
  {author} {\bibfnamefont {V.}~\bibnamefont {Gheorghiu}}, \ and\ \bibinfo
  {author} {\bibfnamefont {R.~B.}\ \bibnamefont {Griffiths}},\ }\href {\doibase
  10.1103/PhysRevA.78.042303} {\bibfield  {journal} {\bibinfo  {journal} {Phys.
  Rev. A}\ }\textbf {\bibinfo {volume} {78}},\ \bibinfo {pages} {042303}
  (\bibinfo {year} {2008})}\BibitemShut {NoStop}%
\bibitem [{\citenamefont {Keet}\ \emph {et~al.}(2010)\citenamefont {Keet},
  \citenamefont {Fortescue}, \citenamefont {Markham},\ and\ \citenamefont
  {Sanders}}]{keetQuantumSecretSharing2010}%
  \BibitemOpen
  \bibfield  {author} {\bibinfo {author} {\bibfnamefont {A.}~\bibnamefont
  {Keet}}, \bibinfo {author} {\bibfnamefont {B.}~\bibnamefont {Fortescue}},
  \bibinfo {author} {\bibfnamefont {D.}~\bibnamefont {Markham}}, \ and\
  \bibinfo {author} {\bibfnamefont {B.~C.}\ \bibnamefont {Sanders}},\ }\href
  {\doibase 10.1103/PhysRevA.82.062315} {\bibfield  {journal} {\bibinfo
  {journal} {Phys. Rev. A}\ }\textbf {\bibinfo {volume} {82}},\ \bibinfo
  {pages} {062315} (\bibinfo {year} {2010})}\BibitemShut {NoStop}%
\bibitem [{\citenamefont
  {Helwig}(2013)}]{helwigAbsolutelyMaximallyEntangled2013}%
  \BibitemOpen
  \bibfield  {author} {\bibinfo {author} {\bibfnamefont {W.}~\bibnamefont
  {Helwig}},\ }\href@noop {} {\enquote {\bibinfo {title} {Absolutely
  {{Maximally Entangled Qudit Graph States}}},}\ } (\bibinfo {year} {2013}),\
  \Eprint {http://arxiv.org/abs/1306.2879} {arXiv:1306.2879} \BibitemShut
  {NoStop}%
\bibitem [{\citenamefont {Rossi}\ \emph {et~al.}(2013)\citenamefont {Rossi},
  \citenamefont {Huber}, \citenamefont {Bru{\ss}},\ and\ \citenamefont
  {Macchiavello}}]{rossiQuantumHypergraphStates2013}%
  \BibitemOpen
  \bibfield  {author} {\bibinfo {author} {\bibfnamefont {M.}~\bibnamefont
  {Rossi}}, \bibinfo {author} {\bibfnamefont {M.}~\bibnamefont {Huber}},
  \bibinfo {author} {\bibfnamefont {D.}~\bibnamefont {Bru{\ss}}}, \ and\
  \bibinfo {author} {\bibfnamefont {C.}~\bibnamefont {Macchiavello}},\ }\href
  {\doibase 10.1088/1367-2630/15/11/113022} {\bibfield  {journal} {\bibinfo
  {journal} {New J. Phys.}\ }\textbf {\bibinfo {volume} {15}},\ \bibinfo
  {pages} {113022} (\bibinfo {year} {2013})}\BibitemShut {NoStop}%
\bibitem [{\citenamefont {Qu}\ \emph {et~al.}(2013)\citenamefont {Qu},
  \citenamefont {Wang}, \citenamefont {Li},\ and\ \citenamefont
  {Bao}}]{quEncodingHypergraphsQuantum2013}%
  \BibitemOpen
  \bibfield  {author} {\bibinfo {author} {\bibfnamefont {R.}~\bibnamefont
  {Qu}}, \bibinfo {author} {\bibfnamefont {J.}~\bibnamefont {Wang}}, \bibinfo
  {author} {\bibfnamefont {Z.-s.}\ \bibnamefont {Li}}, \ and\ \bibinfo {author}
  {\bibfnamefont {Y.-r.}\ \bibnamefont {Bao}},\ }\href {\doibase
  10.1103/PhysRevA.87.022311} {\bibfield  {journal} {\bibinfo  {journal} {Phys.
  Rev. A}\ }\textbf {\bibinfo {volume} {87}},\ \bibinfo {pages} {022311}
  (\bibinfo {year} {2013})}\BibitemShut {NoStop}%
\bibitem [{\citenamefont {Steinhoff}\ \emph {et~al.}(2017)\citenamefont
  {Steinhoff}, \citenamefont {Ritz}, \citenamefont {Miklin},\ and\
  \citenamefont {G{\"u}hne}}]{steinhoffQuditHypergraphStates2017}%
  \BibitemOpen
  \bibfield  {author} {\bibinfo {author} {\bibfnamefont {F.~E.~S.}\
  \bibnamefont {Steinhoff}}, \bibinfo {author} {\bibfnamefont {C.}~\bibnamefont
  {Ritz}}, \bibinfo {author} {\bibfnamefont {N.~I.}\ \bibnamefont {Miklin}}, \
  and\ \bibinfo {author} {\bibfnamefont {O.}~\bibnamefont {G{\"u}hne}},\ }\href
  {\doibase 10.1103/PhysRevA.95.052340} {\bibfield  {journal} {\bibinfo
  {journal} {Phys. Rev. A}\ }\textbf {\bibinfo {volume} {95}},\ \bibinfo
  {pages} {052340} (\bibinfo {year} {2017})}\BibitemShut {NoStop}%
\end{thebibliography}
\end{document}